\documentclass[11pt]{article}
\pdfoutput=1
\usepackage{amsmath}
\usepackage{amssymb}
\usepackage{multirow}

\usepackage{graphicx,bbm,mathrsfs}
\usepackage{nicefrac}
\usepackage{bbm}
\usepackage{geometry}       
\usepackage{jheppub}     
\usepackage{bm}        
\usepackage{graphicx,mathrsfs}
\usepackage{nicefrac}
\usepackage{color}
\definecolor{orange}{RGB}{ 148,0,211}

\def\ie{{\it i.e.}}

\newcommand{\be}{\begin{equation}}  
\newcommand{\ee}{\end{equation}}  
\newcommand{\bea}{\begin{eqnarray}}  
\newcommand{\eea}{\end{eqnarray}}  

\newcommand{\tr}{\operatorname{tr}}

\newcommand{\bmat}{\begin{pmatrix}}
\newcommand{\emat}{\end{pmatrix}}

\newcommand*{\Scale}[2][4]{\scalebox{#1}{$#2$}}%

\newcommand{\nn}{\nonumber}

\begin{document}

\vspace*{1.2cm}

\begin{center}

\thispagestyle{empty}
{\Large\bf 
Sharpening the shape analysis for  higher-dimensional operator searches
}\\[10mm]

\renewcommand{\thefootnote}{\fnsymbol{footnote}}

{\large  Sylvain~Fichet$^{\,a\,}$,  
Patricia Rebello Teles $^{\,b,\,c\,}$, 
Alberto Tonero$^{a\,}$
\footnote{sylvain@ift.unesp.br, patricia.rebello.teles@cern.ch, alberto.tonero@gmail.com   } }\\[10mm]

\addtocounter{footnote}{-1}

{\it $^a$ ICTP-SAIFR \& IFT-UNESP, R. Dr. Bento Teobaldo Ferraz 271, S\~ao Paulo, Brazil \\
}
{ \it $^b$ CERN, CH-1211 Geneva 23, Switzerland \\
 }
{ \it $^c$ CBPF, R. Dr. Xavier Sigaud 150, Rio de Janeiro, Brazil\\
 }

\vspace*{12mm}

{  \bf  Abstract }
\end{center}

When the Standard Model is interpreted as the renormalizable sector of a low-energy effective theory, the effects of new physics  are encoded into a set of higher dimensional operators. 
These operators 
 potentially deform the shapes of Standard Model differential distributions of final states observable at colliders. 
 We describe a simple and systematic method to obtain optimal estimations of these deformations when using numerical tools, like Monte Carlo simulations. A crucial aspect of this method is  minimization of the estimation uncertainty: we demonstrate how the operator coefficients have to be set in the simulations in order to get optimal results. 
  The uncertainty on the interference term turns out to be the most difficult to control and grows very quickly when the interference is suppressed. 
   We exemplify our method by computing the deformations induced by the ${\cal O}_{3W}$ operator in $W^+W^-$ production at the LHC, and by deriving a bound on ${\cal O}_{3W}$ using $8$~TeV CMS data.

\noindent
\clearpage

\section{Introduction}\label{se:intro}

After the historical discovery of the Higgs boson~\cite{cms_higgs,atlas_higgs} in July 2012, the Large Hadron Collider (LHC) is currently probing matter and spacetime at unprecedented small distances, looking for  a signal of  physics beyond the Standard Model (SM). However, new physics remains very elusive so far, as no statistically significant signal has been observed in the collected data up to now.
Although there are good theoretical arguments to expect new particles nearby the TeV scale, it is also  plausible that these states be somewhat too heavy to be on-shell produced at the LHC. In such scenario, the presence of new physics states is best studied using effective field theory methods and their effects can be parametrized by higher dimensional operators made of SM fields. The search for heavy  physics beyond the SM then becomes a program of SM precision measurements, aiming at testing the existence of one or several of these higher dimensional effective operators.

The analysis of the  distributions of  final states kinematic variables  plays a central role in this scenario, because their shapes  contain important information about the presence of the effective operators.  
 In principle, a thorough analysis of the shapes of the differential distributions may be the key to the discovery of new physics. However, such program is not so straightforward to carry out systematically, because of the large number of effective operators and kinematic variables to take into account, and  because of the computational cost of estimating the differential rates. One-dimensional differential rates made available by ATLAS and CMS in run-I Higgs analyses have been included in the  global fits performed in Refs.~\cite{Ellis:2014jta,Corbett:2015ksa,Butter:2016cvz}.  Many progresses are still needed in order to systematically and efficiently search for an arbitrary number of higher dimensional operators using multidimensional differential rates. An attempt to improve  shape analysis techniques using the moments of the differential distributions has been done in~\cite{Fichet:2014fxa}. 
 
From the theoretical point of view, it is crucial to be able to optimize the determination of the various contributions (SM, interference and BSM) that constitute the differential rates in presence of dimension-6 operators, especially when a numerical estimator (\textit{e.g.} Monte-Carlo tool) is available . In the \texttt{MadGraph5} framework~\cite{Alwall:2011uj}, it has become possible to estimate separately the various components of the event rates induced by effective operators, however this useful improvement is not, for the time being, applicable to any kind of processes. It does not apply when several processes are combined, for example when asking for production of a  particle on the mass-shell followed by its decay. Also, if the effective operators enter in the propagators, or if the energy dependence of the width has to be taken into account, the option cannot be consistently used. 

Whenever we face the situation in which this \texttt{MadGraph5} option to separate the various components cannot be used, or another event generator is used, an alternative method has to be invoked for the optimal determination of the contributions (SM, interference and BSM) that constitute the differential distributions. 
The derivation of such method, valid in general, is the topic of this work. 

This paper is organized as follows. The inspection of the event rates in presence of effective operators is done in section~\ref{se:phasespace}.  Based on these considerations, an optimal method to estimate the deformations of differential rates induced by an arbitrary number of effective operators  is presented in section~\ref{se:shape_analysis_EFT}. As illustration and check, in section~\ref{se:searches}, we obtain the deformations induced by the ${\cal O}_{3W}$ operator to the $WW$ production differential rate and we reproduce, within two sigma, the bound obtained by CMS on the coefficient of the ${\cal O}_{3W}$ operator using $8$~TeV data . Conclusions are presented in section~\ref{se:conc}.

\section{Low-energy amplitudes and phase-space considerations}
\label{se:phasespace}

\subsection{Effective theory basics}

A field theory can sometimes be approximated by a simpler, so-called ”effective” theory, in a given region of  phase space.~\footnote{ One often encounters low-energy effective theories in the literature,  but there are other regions of phase space where an effective theory approach can be used. For example, an effective theory for singlet resonances has been described in \cite{Fichet:2015yia}.} 
Consider an ultraviolet (UV) theory that lies at a scale  $\Lambda$ which is much larger than the typical energy scale $E$ of a given experiment $(\Lambda\gg E)$. From the UV point of view, the \textit{low-energy} effective theory is obtained by expanding the correlation functions computed within the UV theory with respect to the parameter $E/\Lambda$.

It is very plausible that the Standard Model is not the ultimate UV-complete theory of nature and its Lagrangian corresponds to the relevant and marginal sector of a low-energy effective 
theory. 
In this picture, the full SM effective Lagrangian also contains a series of operators of dimension higher than four, built from SM fields and invariant under SM symmetries, which are suppressed by inverse powers of the new physics scale $\Lambda$. To the low-energy observer, these higher dimensional operators parametrize the new physics effects, which appear as new interactions between the known particles, including vertices with extra derivatives. The SM effective Lagrangian takes the general form
\be
	\mathcal{L}_{\rm eff} ~=~ \mathcal{L}^{\rm SM} + \sum_{d=5}^\infty \mathcal{L}^{(d)}  \,,
\ee
where
\be 
	\mathcal{L}^{(d)}= \sum_{I}\frac{\alpha_{I}^{(d)}}{\Lambda^{d-4}}\mathcal{O}_I^{(d)}\,.
\ee
The effective operators $\mathcal{O}_I^{(d)}$ have mass dimension $d>4$ and 
$\alpha_{I}^{(d)}$ are dimensionless coefficients. In general, these coefficients can be complex, depending on the operator. However one can always split the operator into its real and imaginary parts in order to end up with only real coefficients. Because of this, in the following we will consider only real coefficients, without loss of generality.

Identifying $\Lambda$ with the mass scale of a heavy particle, the effective field theory is  valid only at energies
\be 
E<\Lambda
\label{eq:EFT_val}
\ee
and the contributions of the effective operators to any given observable are arranged as an expansion in $E/\Lambda$.~\footnote{Another bound comes from requiring a perturbative expansion of the effective interactions, which implies
\be
\frac{|\alpha|}{\Lambda^2}< \frac{4\pi }{E^2}\,.
\ee
Moreover, requiring perturbative couplings in the UV theory implies a bound on $\alpha$, that depends on the number of fields $n$ in the effective operator (or equivalently to the number of legs of the corresponding UV amplitudes). For dimension-6 operators  one has
\be
|\alpha|<(4\pi)^{n-2}\,,
\ee
with $n\leq 6$.   More details about the validity of the SM effective field theory in relation to possible UV completions can be found in \cite{Contino:2016jqw}.
}
The power of the effective theory approach lies in the fact that this series can be consistently truncated in order to match a desired precision, which can be chosen accordingly to the available experimental uncertainties. At any given order of truncation, new physics effects are described by a finite set of universal coefficients. Larger effects are expected to come from operators with lower dimension: for example, the leading contributions to Higgs observables come from the dimension-6 Lagrangian $\mathcal{L}^{(6)}$, while the leading contributions to quartic neutral gauge boson interactions come from the dimension-8 Lagrangian $\mathcal{L}^{(8)}$.~\footnote{There can sometimes be selection rules suppressing the interference of certain dimension-6 operators with the SM \cite{Azatov:2016sqh}. In that case dimension-6 and 8 operators have comparable impact. Just in case, we stress that such feature has nothing to do with a possible breakdown of the $E/\Lambda$ expansion. The effects of operators with $d>8$ are still expected to be suppressed by powers of $E/\Lambda$ with respect to the leading effects. Possible selection rules on the $d>8$ operators would only increase even more this suppression.}
 Here and in the following we will truncate the EFT expansion at dimension 6. The complete list of independent dimension-6 SM effective operators have been reported in \cite{Grzadkowski:2010es}.

Let us consider, for concreteness, the case of a single  dimension-6 effective operator. The general case involving an arbitrary number of higher dimensional operators is very similar. The effective lagrangian is then given by  
\be
\mathcal{L}_{\rm eff} ~=~ \mathcal{L}^{\rm SM} +\frac{\alpha}{\Lambda^{2}}\mathcal{O}^{(6)}\,.\label{eq:L6}
\ee
In the presence of the effective operator of Eq.~\eqref{eq:L6}, a generic amplitude $\mathcal{M}$ can be expressed  as  
\be
\label{eq:M_initial}
\mathcal{M}=\mathcal{M}^{\rm SM}+\frac{\alpha}{\Lambda^2} \mathcal{M}^{\rm BSM}+O\left(\frac{\alpha^2}{\Lambda^4}\right)\,,
\ee 
where $\mathcal{M}^{\rm SM}$ is the SM piece and $\mathcal{M}^{\rm BSM}$ represents the leading BSM contribution obtained by one insertion of the effective operator. The subleading contributions are given by Feynman diagrams with more than one insertion of the effective operator. 
 Notice that, in general, the BSM component $\mathcal{M}^{\rm BSM}$ is not proportional to $ \mathcal{M}^{\rm SM}$. 
\begin{figure}
\centering
\begin{picture}(300,60)
\put(0,-10){        \includegraphics[trim=0cm 0cm 0cm 0cm, clip=true,width=10cm]{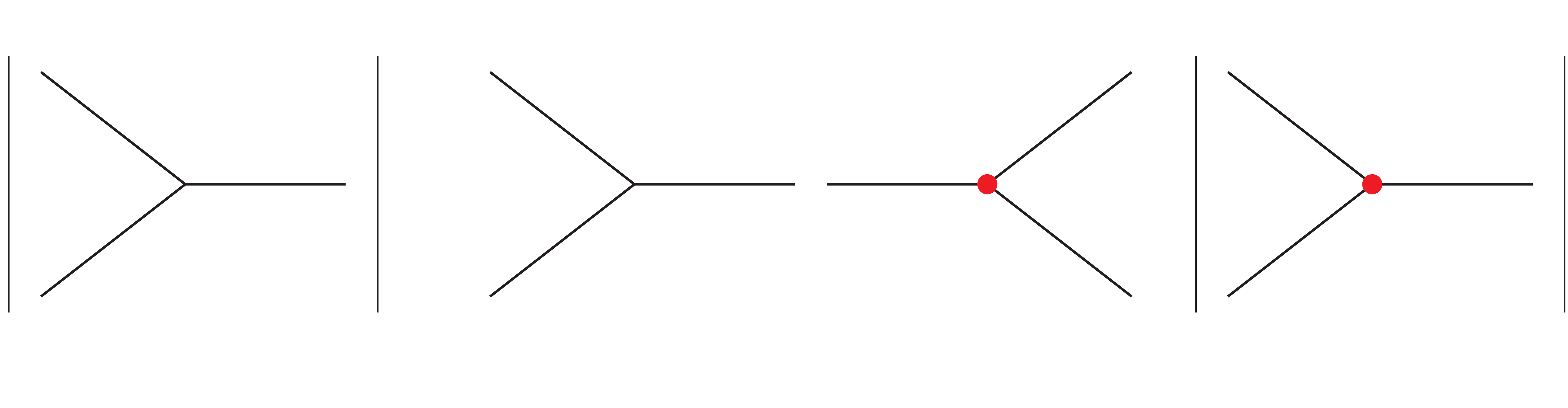}}
\put(290,50){$2$}
\put(75,50){$2$}
\put(85,27){$+$}
\put(208,27){$+$}
\put(230,-20){$\alpha^⁴\Lambda^{-4}\sigma^{\rm BSM}$}
\put(130,-20){$\alpha\Lambda^{-2}\sigma^{\rm int}$}
\put(30,-20){$\sigma^{\rm SM}$}
\end{picture}
\vspace{0.5cm}
    \caption{ The leading contributions to the square amplitude in presence of a dimension-6 operator.
    \label{fig:ME_EFT}}
    \end{figure}
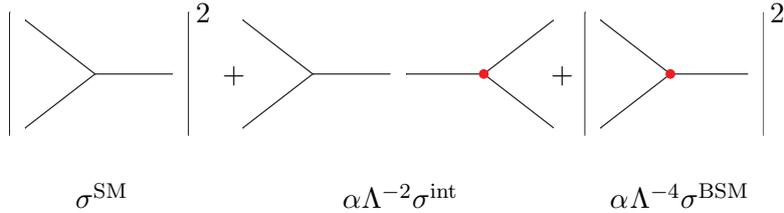
    
Physical observables, like the ones measured at the LHC, are described statistically by event rates $\sigma$ that are given by the integral of the squared amplitude $|\mathcal M|^2$ 
over some phase space domain $\mathcal{D}$ --which can be chosen by the experimentalist to some extent. Typical examples of such observables are total cross-sections and decay widths. 
In the presence of the effective operator of Eq.~\eqref{eq:L6}, the observable event rate $\sigma$ is given by  \footnote{In the following we will refer to differential cross sections as event rates. We remind that, strictly speaking,  an event rate is rather defined as  a cross section times the instantaneous luminosity of the collision. }
\be \label{rates}
\sigma \propto \int_{\mathcal D}\,d\Phi \, \left| \mathcal{M} \right|^2\,,
\ee
where $\mathcal{M}$ is given by Eq.~\eqref{eq:M_initial} and $\int_{\mathcal D} d\Phi$ denotes the integral over the phase space domain.
This event rate $\sigma$ can be further decomposed as a sum of three leading contributions
\be\sigma=\sigma(\alpha)\equiv\sigma^{\rm SM}+\frac{\alpha}{\Lambda^2}\,\sigma^{\rm int}+\frac{\alpha^2}{\Lambda^4}\,\sigma^{\rm BSM}+\ldots\,,\label{eq:rate}
\ee 
where 
\be
\sigma^{\rm SM} \propto \int_{\mathcal D}\,d\Phi \, \left| \mathcal{M}^{\rm SM}  \right|^2
\,, \quad
\sigma^{\rm int} \propto \int_{\mathcal D}\,d\Phi \, 2{\rm Re}\left[\mathcal{M}^{\rm BSM} \mathcal{M}^{\rm SM*}\right]
\,, \quad
\sigma^{\rm BSM} \propto \int_{\mathcal D}\,d\Phi \, \left| \mathcal{M}^{\rm BSM}  \right|^2
\,.
\ee
The ellipses in Eq.~\eqref{eq:rate} represent both contributions of higher order in $1/\Lambda$,  and $O(\Lambda^{-4})$ terms coming from the interference of higher-order diagrams with the SM component. It will be made clear in next subsection  why all the terms in Eq.~\eqref{eq:rate} should be kept in order to describe the leading effects of new physics.

The leading components of a square matrix element in terms of Feynman diagrams is illustrated in Fig.~\ref{fig:ME_EFT}. 
The component $\sigma^{\rm SM}$ corresponds to the pure SM contribution, which remains in the limit $\alpha/\Lambda^2\rightarrow 0$. The component $\sigma^{\rm int}$ is obtained from the interference between the SM and BSM amplitudes, while the  component $\sigma^{\rm BSM}$ is the pure BSM contribution. 
From the point of view of a new physics searcher, these two latter components together constitute the new physics signal $\sigma^{\rm NP}=\sigma-\sigma^{\rm SM}$, while $\sigma^{\rm SM}$ constitutes the irreducible background.~\footnote{A necessary condition for the shape analysis of a differential distribution to 
provide  information on new physics is that $\mathcal{M}^{\rm BSM}$ be not  proportional  $\mathcal{M}^{\rm SM}$. In practice, this condition is realized most of the time. }
Notice that, by definition,  $\sigma^{\rm int}$ and $\sigma^{\rm BSM}$ have not the same dimension as $\sigma^{\rm SM}$.

Finally, we  notice the general fact that the modulus of the  interference term has an upper bound
\be
|\sigma^{\rm int}|< 2\sqrt{\sigma^{\rm SM}}\sqrt{\sigma^{\rm BSM}}\,.\label{eq:int_bound}
\ee
This  is obtained using
\be
\left| \int d\Phi\,
{\rm Re} [\mathcal{M}^{\rm SM} \mathcal{M}^{{\rm BSM}*}]\right|\leq \left|\int d\Phi\, \mathcal{M}^{\rm SM} \mathcal{M}^{{\rm BSM}*}\right|
\leq \sqrt{\int d\Phi\,|\mathcal{M}^{\rm SM}|^2}\sqrt{\int d\Phi|\mathcal{M}^{{\rm BSM}}|^2}\,,\label{eq:CS}
\ee
 where last step we have used the Cauchy-Schwartz inequality.  In the following, we will refer to Eq.~\eqref{eq:int_bound} as  
  the ``Cauchy-Schwartz bound" on the interference.

\subsection{On the consistent truncation of event rates}

We can now consider various limit cases of signal and background over a given phase space domain.
Consider first a region of phase space  $\mathcal{D}_1$ where the SM contribution is much larger than the BSM contribution
\be
\left|\mathcal{M}^{\rm SM}\right|\gg \left|\frac{\alpha}{\Lambda^2}\mathcal{M}^{\rm BSM}\right|\,.\label{eq:D1}
\ee
The Cauchy-Schwartz bound automatically implies  that the interference 
is much smaller than $\sigma^{\rm SM}$, namely $\sigma^{\rm SM}\gg |\alpha\Lambda^{-2}\sigma^{\rm int}|$.
The modulus of the interference term can in principle take any value between zero and $2|\alpha|\Lambda^{-2}\sqrt{\sigma^{\rm SM}\sigma^{\rm BSM}}$. If we assume the case of an unsuppressed interference then we have 
\be 
\sigma_{\rm SM}\gg\left|\frac{\alpha}{\Lambda^2}\sigma^{\rm int}\right|\gg\frac{\alpha^2}{\Lambda^4}\sigma^{\rm BSM}\,. \label{eq:caseD1}
\ee
In this case, the SM background is large with respect to the signal, and the signal appears predominantly through the interference term $\sigma^{\rm int}$.
The component $\sigma^{\rm BSM}$ can thus be neglected at leading order in the effective theory expansion.
For practical purposes, one may notice that the $\sigma^{\rm BSM}$ term can also be kept, as long as it is  negligible. 
Whenever the $\sigma^{\rm BSM}$ term becomes non-negligible,  the truncation of the EFT expansion has to be pushed to next-to-leading order: The higher order diagrams in Eq.~\eqref{eq:M_initial} have to be taken into account, as well as the contributions from higher-order operators. 
In this paper our focus is merely  on the leading effects new physics, thus aspects of the EFT at next-to-leading order are beyond our scope.  

Still in the case described by Eq.~\eqref{eq:D1}, if the interference is vanishing, then the signal comes mainly from $\sigma^{\rm BSM}$. In such case, one should be careful with the next-to-leading order contributions of the EFT, as described in the paragraph above. The $\sigma^{\rm BSM}$ component being $O(1/\Lambda^4)$, other $O(1/\Lambda^4)$ contributions coming from the interference of higher order diagrams or dimension-8 operators with the SM amplitude can be present. These observations have also been made in \cite{Contino:2016jqw,Azatov:2016sqh}.

Consider now a phase space domain $\mathcal{D}_2$ where the SM contribution is much smaller than the BSM contribution
\be
\left|\mathcal{M}^{\rm SM}\right|\ll \left|\frac{\alpha}{\Lambda^2}\mathcal{M}^{\rm BSM}\right|\label{eq:D2}\,.
\ee 
Using again the Cauchy-Schwartz bound, Eq.~\eqref{eq:D2} automatically implies that the interference term is much smaller than the BSM contribution, namely $\alpha^2\Lambda^{-4}\sigma^{\rm BSM}\gg |\alpha\Lambda^{-2}\sigma^{\rm int}|$.
Assuming  an unsuppressed interference,  we have 
\be 
\sigma^{\rm SM}\ll\left|\frac{\alpha}{\Lambda^2}\sigma^{\rm int}\right|\ll\frac{\alpha^2}{\Lambda^4}\sigma^{\rm BSM}\,.
\ee
This is the typical situation of a search for ``rare events'', 
where the SM background is vanishing and each signal event carries a high statistical significance.
 In this case, the new physics signal appears predominantly through the pure BSM term $\sigma^{\rm BSM}$. This implies that, even at leading order in the EFT expansion, one has to keep the $\sigma^{\rm BSM}$ term. This fact might seem puzzling at first view, as one ends up with a $O(1/\Lambda^4)$ term at leading order.  
Naively, there are other $O(1/\Lambda^4)$ contributions coming from the interference of higher order diagrams or dimension-8 operators with the SM amplitude. 
However, one can easily check that the contribution $\sigma^{\rm BSM}$ is actually the dominant one, because the other $O(\Lambda^{-4})$ contributions would  come from an interference with the SM amplitude, which is small by assumption (see Eq.~\eqref{eq:D2}).

We conclude from the above analysis  that in general, \textit{both} the interference term $\sigma^{\rm int}$ and the quadratic BSM term $\sigma^{\rm BSM}$ have to be kept in order to describe the leading effects of new physics for any background configuration. The 
$\sigma^{\rm int}$ piece dominates when $\left|\mathcal{M}^{\rm SM}\right|\gg \left|\alpha\Lambda^{-2}\mathcal{M}^{\rm BSM}\right|$ --provided that the interference is not suppressed--while the $\sigma^{\rm BSM}$ piece dominates for $\left|\mathcal{M}^{\rm SM}\right|\ll \left|\alpha\Lambda^{-2}\mathcal{M}^{\rm BSM}\right|$.

Finally, let us comment about specific event rates encountered in collider experiment. In the case of resonant production of an unstable particle, the propagator is resummed, so that the amplitude has not initially  the form  Eq.~\eqref{eq:M_initial}, but the form
\be
|\mathcal{M}|^2_{\rm res}\propto\frac{1}{(s-m^2)^2+m^2\Gamma_{\rm tot}^2}\,,
\ee
where $\Gamma_{\rm tot}$ is the total decay width. 
This is instead the total width which has the form of a quadratic function, $\Gamma_{\rm tot }=\Gamma^{\rm SM}+\alpha \Lambda^{-2} \Gamma^{\rm int} +\alpha^2 \Lambda^{-4} \Gamma^{\rm BSM}$. However, if the EFT expansion is valid, one can always expand $|\mathcal{M}|^2_{\rm res}$ with respect to $s/\Lambda^2$ and $m^2/\Lambda^2$, to end up with a quadratic form    $|\mathcal{M}|^2_{\rm res, SM}+\alpha \Lambda^{-2} |\mathcal{M}|^2_{\rm res, int}+\alpha^2 \Lambda^{-4} |\mathcal{M}|^2_{\rm res, BSM}+O(\Lambda^{-6})$.
Similarly, when using the narrow-width approximation in a production+decay process, the event rates take the form $\sigma_i\equiv\sigma_{\rm prod} \Gamma_i/\Gamma_{\rm tot}$, where $\sigma_{\rm prod}$ is a production cross-section and $\Gamma_i$ a partial decay width. Again, the three quantities $\sigma_{\rm prod}$, $\Gamma_i$, $\Gamma_{\rm tot}$ are in principle quadratic functions of $\alpha$. However, if the EFT expansion is valid, one can always expand $\sigma_i$ to quadratic order so that $\sigma_i=\sigma^{\rm SM}+\alpha\Lambda^{-2}\sigma^{\rm int}+\alpha^2\Lambda^{-4}\sigma^{\rm BSM}+O(\Lambda^{-6})$.

We can see that the EFT expansion always allows us to express the event rate as a quadratic function in $\alpha/ \Lambda^2$, even in case of resonances.  The discussion of this section applies similarly to the case of an arbitrary number of effective operators.

\paragraph{A comment on the relevance of rare events}

One may remark that in the case where we have $\left|\mathcal{M}^{\rm SM}\right|\gg \left|\alpha\Lambda^{-2}\mathcal{M}^{\rm BSM}\right|$ and an unsuppressed interference, 
the signal is  proportional to $\alpha/\Lambda^2$. On the other hand, in the case $\left|\mathcal{M}^{\rm SM}\right|\ll \left|\alpha\Lambda^{-2}\mathcal{M}^{\rm BSM}\right|$, the signal
is proportional to $\alpha^2/\Lambda^4$. The absolute magnitude of the signal is thus much larger in the former case than in the latter one, but the magnitude of the \textit{background} is also larger in the former case. We may therefore wonder which of these two configurations is the more advantageous in order to detect the signal.  The answer to this seemingly straightforward question is not so easy, and needs to involve a test statistic.

Assume a signal discovery test whose significance $Z$ is given by~\footnote{This is the significance from the  p-value of a likelihood-ratio test, assuming a counting experiment with large data sample (see e.g.~\cite{Cowan:2010js}). }
 \be
Z=\frac{N-N_{\rm bkg}}{\sqrt{N_{\rm bkg}}}\,,\label{eq:Z_disc}
\ee
where $N$ is the total number of events and $N_{\rm bkg}$ is the expected number of background events.
Let $\mathcal{D}_1$ and $\mathcal{D}_2$ be two domains of phase space satisfying
\be\sigma^{\rm BSM}_{ 1}\approx \sigma^{\rm BSM}_{ 2}\,,\quad\sigma^{\rm SM}_{ 1}\gg \left|\frac{\alpha}{\Lambda^2}\sigma^{\rm int}_{ 1}\right|\gg \frac{\alpha^2}{\Lambda^4}\sigma^{\rm BSM}_{ 1} \,,\quad\sigma^{\rm SM}_{ 2}\ll \left|\frac{\alpha}{\Lambda^2}\sigma^{\rm int}_{ 2}\right|\ll \frac{\alpha^2}{\Lambda^4}\sigma^{\rm BSM}_{ 2} \label{eq:ass} \,,\ee
where $\sigma_{ 1,2}^{\rm SM}$, $\sigma^{\rm int}_{ 1,2}$ and $\sigma^{\rm BSM}_{ 1,2}$ are the components of the event rate $\sigma$ on $\mathcal{D}_{1,2}$.
Denoting $Z_{1}$ and $Z_{2}$  the  discovery significances on $\mathcal{D}_1$ and $\mathcal{D}_2$ respectively, we have
\be
\frac{Z_1}{Z_2} \ll 2 \,. \label{eq:res}
\ee
The proof is given in App. \ref{app:proof_Z}.
This small theorem makes clear that the regions of phase space where 
$|\mathcal{M}^{\rm SM}|\ll |\alpha\Lambda^{-2}\mathcal{M}^{\rm BSM}|$, \ie~``rare events'' regions, can provide a lot of statistical significance, even though the signal is much weaker in that region compared to the 
$|\mathcal{M}^{\rm SM}|\gg |\alpha\Lambda^{-2}\mathcal{M}^{\rm BSM}|$ 
region. The signal search in such ``rare events'' regions deserves thus a particular attention. We further emphasize that this ``rare events'' situation naturally happens whenever a higher dimensional operator carries derivatives, as is the case for many of the SM dimension-6 operators. For such operators, the high-energy tails of the kinematic distributions are typically ``rare events'' zones. The analysis of high-energy tails deserves thus particular attention. For example, regarding presentation of results, the use of overflow bins for the high-energy tail should be prevented as much as possible in order not to lose the precious information from the tail. Instead, all the bins should be kept up to the highest energetic  event, no matter how few events are contained in the bins.

\section{Optimal estimation of the differential rates}
\label{se:shape_analysis_EFT}

In the previous section we have considered event rates over a region of phase space $\mathcal{D}$. Let us now assume that the experiment allows to partition the region $\mathcal{D}$ into subdomains such that $\mathcal{D}=\cup_r \mathcal{D}_r$. We refer to the  $\mathcal{D}_r$ subdomains as ``bins''.  In addition to the total rate, the knowledge of the event rate in each bin provides information about the \textit{shape} of the distribution. From the experimental point of view one talks about binned distribution. In principle, the size of each bin can be made small enough so that each event is seen separately. In this case one talks about unbinned distribution. Notice that, since the experimental precision is finite, the bin size can never shrink to zero, however it is instructive to keep in mind that the latter case can be seen as a limit of the former.

Let us first consider  the case of one single dimension-6 effective operator whose contribution to the amplitudes is given by Eq. (\ref{eq:M_initial}).
Let $X$ be a variable of phase space with domain $\mathcal{D}_X \subset \mathcal{D}$. For a collider experiment $X$ can be some transverse momentum, invariant mass, \ldots One defines the ``differential event rate'' along $X$ as 
\be \label{difrate}
\sigma_X \equiv \frac{d\sigma}{dX}\propto \int_{\mathcal{D}\backslash\mathcal{D}_X}\,d\Phi \, \left| \mathcal{M}^{\rm SM}+\frac{\alpha}{\Lambda^2} \mathcal{M}^{\rm BSM}  \right|^2\,,
\ee 
where $ \int_{\mathcal{D}\backslash \mathcal{D}_X}$ means that the integral is performed over the complement of $\mathcal{D}_X$. 
Integrating $\sigma_X$ over the domain $\mathcal{D}_X$ one recovers the total rate, namely $\int_{\mathcal{D}_X} dX\, \sigma_X\,=\,\sigma$. This $\sigma_X$ should be used when treating unbinned data. The bins $\mathcal{D}_{Xr}$ over the domain $\mathcal{D}_X$ are defined by the partition $\mathcal{D}_X = \cup_r\mathcal{D}_{Xr}$ and the event rate $\sigma_r$ on a bin $\mathcal{D}_{Xr}$ is then given by
\be
\int_{\mathcal{D}_{Xr}} dX\,\sigma_X\,=\,\sigma_r\,.
\ee
The set of the rates on every bin $\{\sigma_{r}\}$ forms an histogram, that constitutes a discrete estimator of the true differential distribution $\sigma_X$.

Let us now proceed to decompose the differential rate $\sigma_X$ defined in Eq. (\ref{difrate}) as a sum of three components $\sigma^{\rm SM}_{X}$, $\sigma^{\rm int}_{X}$ and $\sigma^{\rm BSM}_{X}$, in complete analogy to the case of the total rate $\sigma$. We have
\be\label{difratedec}
\sigma_X=\sigma_X(\alpha)\equiv\sigma^{\rm SM}_{X}+\frac{\alpha}{\Lambda^2}\,\sigma^{\rm int}_{X}+\frac{\alpha^2}{\Lambda^4}\,\sigma^{\rm BSM}_{X}\,,
\ee 
where
\be
\sigma^{\rm SM}_{X} \propto \int_{\mathcal D\backslash\mathcal D_X}\,d\Phi \, \left| \mathcal{M}^{\rm SM}  \right|^2
\,, \ee\be
\sigma^{\rm int}_{X} \propto \int_{\mathcal D\backslash\mathcal D_X}\,d\Phi \, 2{\rm Re}\left[\mathcal{M}^{\rm BSM} \mathcal{M}^{\rm SM*}\right]
\,, \quad
\sigma^{\rm BSM}_{X} \propto \int_{\mathcal D\backslash\mathcal D_X}\,d\Phi \, \left| \mathcal{M}^{\rm BSM}  \right|^2
\,.
\ee
The same is also true for the binned rates
\be\label{binratedec}
\sigma_r=\sigma_r(\alpha)\equiv\sigma^{\rm SM}_{r}+\frac{\alpha}{\Lambda^2}\,\sigma^{\rm int}_{r}+\frac{\alpha^2}{\Lambda^4}\,\sigma^{\rm BSM}_{r}\,.
\ee 

The phase space integral is usually difficult or impossible to evaluate analytically, for example because of the complexity of $\cal D$.  Its evaluation has then to rely on a numerical integration method, for instance a Monte-Carlo simulation. In the following we are going to assume that such estimation method is available.

\subsection{Reconstructing the differential rates}

\label{se:recon}

Assuming we have a way of evaluating a differential rate in presence of effective operators with   coefficients fixed to given values, we can now wonder how to efficiently determine the deformations induced by the effective operators.

We first consider the case of a unique dimension-6  operator.
We have seen in Sec.~\ref{se:phasespace} that the expansion of the event rate has to be kept up to quadratic order in $\alpha$ and the same argument applies also to the differential rate $\sigma_X$. Being $\sigma_X$ a quadratic function of $\alpha$ (see Eq.~\eqref{difratedec}), in principle, it is sufficient to know $\sigma_X$ for only \textit{three} different values of $\alpha$, namely $\alpha_0$, $\alpha_1$ and $\alpha_2$, in order to reconstruct the exact form of $\sigma_X(\alpha)$.
Whenever these three evaluations of $\sigma_X$ are available, that we denote by $\sigma_X^i=\sigma_X(\alpha_i)$, $i=0,1,2$, then the three components $\sigma^{\rm SM}_{X}$, $\sigma^{\rm int}_{X}$ and $\sigma^{\rm BSM}_{X}$ are obtained by just solving a $3\times 3$ linear system and this simple task needs to be carried out only once.

In the case in which the estimations are made by means of Monte Carlo simulations, one directly deals with the binned rates $\sigma_{r}$ which are extracted from the histogram of the $\sigma_X$ distribution. The three components $\sigma^{\rm SM}_{r}$, $\sigma^{\rm int}_{r}$, $\sigma^{\rm BSM}_{r}$ are obtained by solving a $3\times 3$ linear system for each bin. 
The $\sigma^{\rm SM}_{r}$ component can be obtained by simply setting $\alpha=\alpha_0=0$ in the MC simulation. The components $\sigma^{\rm int}_{r}$ and $\sigma^{\rm BSM}_{r}$ are instead obtained by running the MC simulation for \textit{two} non-zero values of $\alpha$, namely $\alpha_1$ and $\alpha_2$. Therefore, we end up with the following solution of our linear system
\bea 
\sigma^{\rm SM}_{r}&=&\sigma_r^0 \nn\\
\sigma^{\rm int}_{r}&=&\frac{\Lambda^2}{\alpha_1\alpha_2}\bigg[
\frac{\alpha_2^2\sigma_{ r}^1-\alpha_1^2\sigma_{r}^1}{\alpha_2-\alpha_1}-(\alpha_1+\alpha_2)\sigma^{\rm 0}_{r}
\bigg] \nn\\
\sigma^{\rm BSM}_{ r}&=&\frac{\Lambda^4}{\alpha_1\alpha_2}\bigg[
-\frac{\alpha_2\sigma_{r}^1-\alpha_1\sigma_{r}^2}{\alpha_2-\alpha_1}+\sigma^{\rm 0}_{r}
\bigg]\,, \label{eq:sys}
\eea
where $\sigma_{r}^0=\sigma_{r}(0)$, $\sigma_{r}^1=\sigma_{r}(\alpha_1)$ and $\sigma_{r}^2=\sigma_{r}(\alpha_2)$.
These  components can then be used in Eq. (\ref{binratedec}), ``reconstructing'' the formula that gives $\sigma_r$ for any value of $\alpha$.

Let us  consider the general case of $n$ effective operators. We present only the case of dimension-6 operators for simplicity. A similar analysis applies to the case of operators with arbitrary dimension.
 The amplitude has in general the form
\be
\mathcal{M}=\mathcal{M}^{\rm SM}+\frac{1}{\Lambda^2}\sum_{I=1}^n \alpha_I \mathcal{M}^{\rm BSM}_{I}+O(\Lambda^{-4})\,.
\ee
The $n$  BSM contributions $\mathcal{M}^{\rm BSM}_{I}$ are in general different one from each other. The differential event rate $\sigma_X$ is proportional to the squared modulus of the amplitude and can be decomposed as 
\be
\sigma_X=\sigma_X(\alpha_I)=\sigma^{\rm SM}_{X}+\frac{1}{\Lambda^2}\sum_{I=1}^n \alpha_I\, \sigma^{\rm int}_{X,I}
+\frac{1}{\Lambda^4}\sum_{I,J=1}^n \alpha_I\alpha_J\, \sigma^{\rm BSM}_{X,IJ}\,.
\ee
It is convenient to rewrite the sum of the BSM quadratic contributions as
\be
\sum_{I,J=1}^n \alpha_I\alpha_J\, \sigma^{\rm BSM}_{X,IJ}=\sum_{I=1}^n \alpha_I^2\, \sigma^{\rm BSM}_{X,II}+2\sum_{I,J=1, I>J}^n \alpha_I\alpha_J\, \sigma^{\rm BSM}_{X,IJ}\label{eq:dec2}
\ee
where $\sigma^{\rm BSM}_{X,IJ}={\rm Re}(\mathcal{M}^{\rm BSM}_{I} \mathcal{M}_{J}^{\rm BSM*})$. The last piece of Eq. (\ref{eq:dec2}) corresponds to the interferences term of the effective operators among themselves. 

In addition to $n$ interference terms $\sigma^{\rm int}_{X,I}$ we have $n$ terms $\sigma^{\rm BSM}_{X,II}$ and $n(n-1)/2$ terms  $\sigma^{\rm BSM}_{X,IJ}$. Including the $\sigma_X^{\rm SM}$ component, the total number of terms to compute is $(n+1)(n+2)/2$. We conclude that $(n+1)(n+2)/2$ simulations are enough to exactly know the event rate $\sigma_X$ as a function of the operators coefficients. The components are obtained by simply solving a $(n+1)(n+2)/2\times (n+1)(n+2)/2$ linear system. This operations needs to be done only once. If one uses histograms, the system has to be solved once for each bin, just like in the case of a single operator.

We conclude that the complexity of the simple reconstruction method describe hereabove grows quadratically with the number of operators, once simplifications provided by the EFT expansion are taken into account.

\subsection{Minimizing the uncertainties } \label{se:Opt}

The numerical evaluations of $\sigma_X$ that are needed to reconstruct $\sigma_X(\alpha)$ are usually endowed with an intrinsic uncertainty. The correct approach is to think about \textit{estimators} of $\sigma_X$ for each value of the chosen $\alpha$. In the case of a single dimension-6 operator the estimators are three and we denote them by $\hat \sigma_X^i=\hat \sigma_X(\alpha_i)$, $i=0,1,2$. 
The uncertainties associated to these estimators are naturally propagated to the $\sigma_X^{\rm SM}$, $\sigma_X^{\rm int}$, $\sigma_X^{\rm BSM}$ components, because these quantities are linear combinations of the  $\sigma_X^i$. In turn, these uncertainties are propagated to the reconstructed $\sigma_X(\alpha)$ distribution because of the relation in Eq.~\eqref{difratedec} and, when it comes to confronting $\sigma_X(\alpha)$ to the observed distribution $\sigma_X^{\rm obs}$, the uncertainty on $\sigma_X(\alpha)$ should be taken into account. It is therefore important to have a measure of this uncertainty, that is given by the \textit{covariance matrix} of the estimators of the $\sigma_X^{\rm SM}$, $\sigma_X^{\rm int}$, $\sigma_X^{\rm BSM}$ components.

In principle, there is a freedom in choosing the $\alpha$ coefficients that are used to perform the numerical estimations of $\sigma_X$ which are needed to reconstruct  $\sigma_X(\alpha)$.
If there were no uncertainties, any set of  values would work fine. However, in presence of uncertainties, it turns out that the choice of the $\alpha$ coefficients has a crucial impact on the reconstruction uncertainties. 
In the following we will determine the values of $\alpha$ that \textit{minimize} these uncertainties.

Let us work with the binned distributions $\sigma_r$ and focus on a bin $r$. The three estimators are denoted by $\hat \sigma_r^0=\hat \sigma_r(\alpha_0)$, $\hat \sigma_r^1=\hat \sigma_r(\alpha_1)$, $\hat \sigma_r^2=\hat \sigma_r(\alpha_2)$. We introduce the relative variance of each estimator $\bar V_r^i$, which is given by 
\be 
\bar V_r^i=\frac{{\rm E}[(\hat \sigma_r^i)^2]-{\rm E}[\hat \sigma_r^i]^2}{{\rm E}[\hat \sigma_r^i]^2}\qquad\qquad(i=0,1,2)\,,
\ee
where $E[\hat y]$ represents the expectation value of the random variable $\hat y$. No correlation is assumed among estimators related to different values of $\alpha_i$.
Furthermore, we assume that the three relative variances have the same magnitude, namely
\be
\bar V_r^0\sim \bar V_r^1\sim \bar V_r^2 \equiv \bar V_r\,. 
\ee
In case the estimators are obtained through  Monte Carlo simulations with $N_{\rm MC}$ points, we have $\bar V_r\sim 1/N_{\rm MC}$. 

\begin{figure}[t]
\centering
\includegraphics[trim=0cm 0cm 0cm 0cm, clip=true,width=10cm]{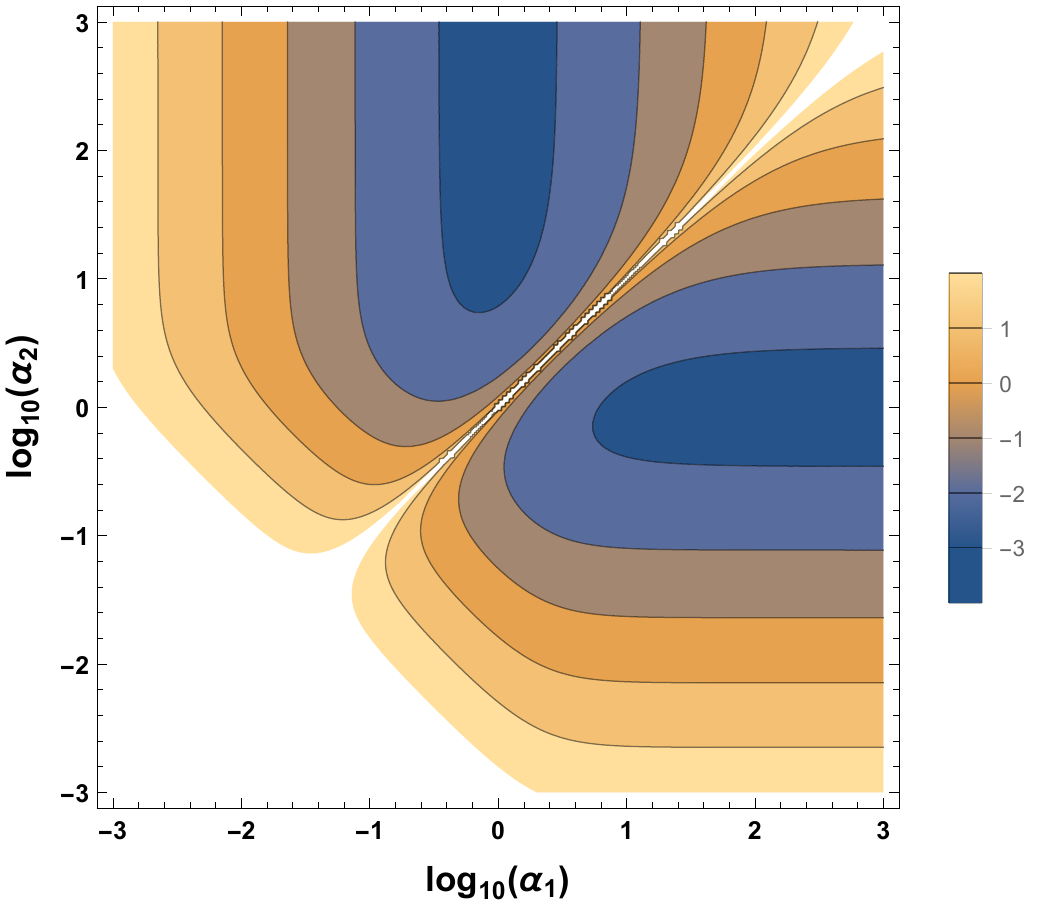}
    \caption{ Trace of the relative covariance of the  $\hat\sigma^{\rm SM}$, $\hat\sigma^{\rm int}$, $\hat \sigma^{\rm BSM}$ estimators as a function of $\alpha_1$, $\alpha_2$. One assumed $\sigma^{\rm SM}\sim\sigma^{\rm BSM}\Lambda^{-4}$, a positive interference saturating the Cauchy-Schwartz bound, and an event number $N_{\rm MC}=10^4$ for each simulation.
    \label{fig:CovOpt}}
    \end{figure}

The estimators of our interest are $\hat \sigma^{\rm SM}_r$, $\hat \sigma^{\rm int}_r$ and $\hat \sigma^{\rm BSM}_r$, which are expressed in terms of some linear combinations of the $\hat \sigma_r^i$ as shown in  Eq.~\eqref{eq:sys}. The quantity at the center of our analysis is the relative covariance matrix of these estimators which is given by
\be
\bar C_r (\alpha_0,\alpha_1,\alpha_2)= 
\begin{pmatrix}
\frac{{\rm E}[(\hat \sigma_r^{\rm SM})^2]}{{\rm E}[\hat \sigma_r^{\rm SM}]^2}-1 &  
\frac{{\rm E}[\hat \sigma_r^{\rm SM} \hat \sigma_r^{\rm int}]}{{\rm E}[\hat \sigma_r^{\rm SM}]{\rm E}[\hat \sigma_r^{\rm int}]}-1 &
\frac{{\rm E}[\hat \sigma_r^{\rm SM} \hat \sigma_r^{\rm BSM}]}{{\rm E}[\hat \sigma_r^{\rm SM}]{\rm E}[\hat \sigma_r^{\rm BSM}]}-1 \\
 &  \frac{{\rm E}[(\hat \sigma_r^{\rm int})^2]}{{\rm E}[\hat \sigma_r^{\rm int}]^2}-1 &
 \frac{{\rm E}[\hat \sigma_r^{\rm int} \hat \sigma_r^{\rm BSM}]}{{\rm E}[\hat \sigma_r^{\rm int}]{\rm E }[\hat \sigma_r^{\rm BSM}]}-1 \\
 & &   \frac{{\rm E}[(\hat \sigma_r^{\rm BSM})^2]}{{\rm E}[\hat \sigma_r^{\rm BSM}]^2}-1
\end{pmatrix}\,.\label{eq:relcov}
\ee
It is important to notice that for the sake of determining the $\sigma_X^{\rm SM}$, $\sigma_X^{\rm int}$ and $\sigma_X^{\rm BSM}$ components, there is \textit{no} need to use values of $\alpha$ and $\Lambda$ that respect the EFT validity bounds of Eq.~\eqref{eq:EFT_val}. Note that to apply this trick,  the possible higher order terms in the expressions of the event rates must be set to zero to avoid any disturbance.~\footnote{We notice that, as the EFT validity bounds do not matter for the sake of determining the components of $\sigma_r(\alpha)$, only the $\alpha/\Lambda^2$ combination actually appears in the problem. A value for $\Lambda$ could thus be fixed without  loss of generality. In the text we will not make such assumption. For Fig.~\ref{fig:CovOpt}, it will be convenient to choose $\Lambda$ so that\be
\sigma^{\rm SM}_r \sim \frac{1}{\Lambda^4}\sigma^{\rm BSM}_r  \,. 
\label{eq:condLambda}
\ee
}

After these preliminary steps, we can discuss the uncertainties that affect the $\hat \sigma^{\rm SM}_r$, $\hat \sigma^{\rm int}_r$ and $\hat \sigma^{\rm BSM}_r$ estimators. We aim at minimizing simultaneously the uncertainties for every component. We should therefore consider the trace of the relative covariance matrix of Eq.~\eqref{eq:relcov}. 
Let us first notice that choosing $\alpha_0=0$ gives simply $\hat \sigma^{\rm SM}_r=\hat \sigma^{0}_r$ and therefore the relative variance for $\hat \sigma^{\rm SM}_r$ is simply $\bar V_r^{\rm SM}=\bar V_r$. This choice is arguably optimal and we are left with finding optimal values for  $\alpha_{1}$ and $\alpha_2$. 

For a fixed value of $\alpha_1$ it turns out that  $\tr \bar C_r(0,\alpha_1,\alpha_2)$ is minimized for $\alpha_2$ going to infinity~\footnote{Because of the  symmetry in ($\alpha_1$, $\alpha_2$) the same is also true for fixed $\alpha_2$ and $\alpha_1\to\infty$.} and this runaway direction can be seen in Fig.~\ref{fig:CovOpt}. In this limit, the $\hat \sigma_r^2$ estimator corresponds exactly to $\hat \sigma^{\rm BSM}_r$ with relative variance equal to $\bar V_r$ and vanishing correlation with $\hat \sigma^{\rm SM}_r$. In this limit the relative covariance matrix takes the form 
\be
\begin{split} & \bar C_r (0,\alpha_1,\infty) =
\\
& \bar V_r 
\begin{pmatrix}
1 & \quad-\frac{\Lambda^2}{\alpha_1}\frac{\sigma_r^{\rm BSM}}{\sigma^{\rm int}_r} & 0 \\
 -\frac{\Lambda^2}{\alpha_1}\frac{\sigma_r^{\rm SM}}{\sigma^{\rm int}_r}\quad &
1+2\frac{ \Lambda^4\sigma^{\rm SM}+\alpha_1^2\sigma_r^{\rm BSM}}{\alpha_1 \Lambda^2 \sigma_r^{\rm int}}
+2\frac{ \Lambda^8(\sigma^{\rm SM})^2+\alpha_1^2 \Lambda^{4}  \sigma^{\rm SM} \sigma_r^{\rm BSM}+
\alpha_1^4 (\sigma_r^{\rm BSM})^2}{\alpha_1^2\Lambda^4 (\sigma_r^{\rm int})^2}
\quad  & -\frac{\alpha_1}{\Lambda^2}\frac{\sigma_r^{\rm BSM}}{\sigma^{\rm int}_r} \\
0 & -\frac{\alpha_1}{\Lambda^2}\frac{\sigma_r^{\rm BSM}}{\sigma^{\rm int}_r}  & 1 
\end{pmatrix}\,. \label{eq:cov_full_gen}
\end{split}
\ee
The trace of the relative variance matrix goes to infinity for both $\alpha_1\rightarrow 0,\infty$ because the linear system becomes degenerate in these limits. Studying further the behaviour of $\tr \bar C_r(0,\alpha_1,\infty)$ we find out that it admits two minima for
\be
\alpha_1 = \pm\Lambda^2\sqrt{\frac{\sigma^{\rm SM}_r}{\sigma^{\rm BSM}_r}}\,.\label{eq:alpha_opt}
\ee
It is a non-trivial feature that this expression is independent on $\sigma_r^{\rm int}$. This implies that the optimization does not depend on whether the interference is suppressed.

The behaviour of the  trace of $\bar C_r(0,\alpha_1,\alpha_2)$ is shown in Fig.~\ref{fig:CovOpt}, where we have taken $\Lambda$ to be such that Eq.~\eqref{eq:condLambda} is verified. One can observe from the plot that the minimum of the trace of the relative covariance matrix occurs for $\alpha_1=1$, $\alpha_2=\infty$ or vice-versa.
The relative covariance matrix at the positive optimal $\alpha_1$ takes the following form  
\be
\bar C_r^{\rm min} =\bar V_r 
\begin{pmatrix}
1 & -\frac{\bar\sigma_r}{\sigma^{\rm int}_r} & 0 \\
 -\frac{\bar\sigma_r}{\sigma^{\rm int}_r} &
1+4\frac{\bar\sigma_r}{\sigma^{\rm int}_r}+6 \left(\frac{\bar\sigma_r}{\sigma^{\rm int}_r}\right)^2  & -\frac{\bar\sigma_r}{\sigma^{\rm int}_r} \\
0 & -\frac{\bar\sigma_r}{\sigma^{\rm int}_r}  & 1 
\end{pmatrix}\,,\label{eq:cov_full_min}\,
\ee
where 
\be
\bar \sigma_r \equiv 
\sqrt{\sigma_r^{\rm SM}\sigma_r^{\rm BSM}}\,.
\ee
Notice that the minimal relative covariance being by definition dimensionless, it can depend on the components of the differential rate only through the combination $\bar \sigma_r/\sigma_r^{\rm int}$. This quantity is also the one appearing in the Cauchy-Schwartz bound of Eq.~\eqref{eq:CS} that now can be rewritten as $|\sigma_r^{\rm int}|\leq 2 \bar \sigma_r$.


Considering the two values of $\sigma_r^{\rm int}$ that saturate the Cauchy-Schwartz bound, the relative covariance matrix at the minimum assumes the following form 
\be
\bar V_r 
\begin{pmatrix}
1 & -\frac{1}{2} & 0 \\
 -\frac{1}{2} &
\frac{9}{2}
  & -\frac{1}{2} \\
0 & -\frac{1}{2}  & 1 
\end{pmatrix}\,\qquad\qquad\mbox{and}\qquad\qquad
\bar V_r 
\begin{pmatrix}
1 &\,\,\, \frac{1}{2}\,\,\,\, & 0 \\
 \frac{1}{2} &
\frac{1}{2}
  & \frac{1}{2} \\
0 & \frac{1}{2}  & 1 
\end{pmatrix}\,, \label{eq:CovOpt}
\ee
for $\sigma^{\rm int}_r=+2 \bar \sigma_r$ and $\sigma^{\rm int}_r=- 2 \bar \sigma_r$ respectively. The relative uncertainty on the interference component is given by the square root of  $(\bar C_r)_{22}$. 
From Eq.~\eqref{eq:CovOpt} one sees that the uncertainty on the interference component is  larger than the ones on $\sigma^{\rm SM}_r$, $\sigma^{\rm BSM}_r$ by at least a  factor $3/\sqrt{2}$ when $\sigma^{\rm int}_r$ is positive. In case of a  maximal negative interference, the optimal $\alpha_1$ corresponds in fact to a fully destructive interference, $\sigma_r^1=0$, and the expected uncertainty associated to this result is vanishing. This can be directly seen in the covariance matrix (Eq.~\eqref{eq:CovOpt}, right side), which features a zero eigenvalue. This situation of a large negative interference gives the smallest uncertainties possible. 
 In contrast, whenever the interference is suppressed, $|\sigma^{\rm int}_r|<2\bar\sigma_r$, the uncertainty  on the interference component quickly blows up. For example, taking $\sigma^{\rm int}_r =\pm 0.2\bar\sigma_r$, we get $(\bar C_r)_{22}\approx 175$ and $(\bar C_r)_{22}\approx 130$ respectively. Our calculation provides a quantitative estimate of the difficulties that will be encountered for determining the interference component.

If one evaluates the minimal relative covariance matrix at the negative optimal $\alpha_1$ of Eq.~\eqref{eq:alpha_opt}, Eq.~\eqref{eq:CovOpt} and the subsequent expressions and discussions are valid up to a sign flip of $\sigma^{\rm int}_r$. Let us notice that if the sign of  $\sigma^{\rm int}_r$ is known in advance,  the sign of the optimal $\alpha_1$ can then be chosen so that the interference term is negative, which gives a better uncertainty  on the estimation of $\sigma^{\rm int}_r$. But this refinement matters mostly for an interference near the Cauchy-Schwartz bound.

Finally, one may notice that the optimization is in principle different for each bin,  as made clear by the $r$-dependence of the optimal point given in  Eq.~\eqref{eq:alpha_opt}. However a simpler version of the optimization can also be obtained by using the $\alpha_1$ that minimizes the uncertainty on the total rates, namely $\alpha_1=\Lambda^2 (\sigma^{\rm SM} / \sigma^{\rm BSM})^{1/2} $.

\subsection{Case of $n$ operators} \label{se:summary}
Let us discuss the reconstruction method in the case of $n$ effective operators.
We have seen in section~\ref{se:recon} that $(n+1)(n+2)/2$ simulations have to be performed to fully reconstruct the event rate $\sigma_X$ as function of the  $\alpha_I$'s. In practice, it turns out it is sufficient to switch on one or two effective operators at a time in order to reconstruct all the components.  
Using the results of the previous section, we can provide the complete list of $(n+1)(n+2)/2$ optimized points to be used for the simulations. We have:
\begin{enumerate}
\item[1)] 
One point with all $\alpha_I=0$. 
\item[2)]
For every $I$,  a  single non-zero $\alpha_I$  satisfying   $\alpha_I\gg 1$. These are $n$ points. 
\item[3)]
For every $I$,  a  single non-zero $\alpha_I$  satisfying   $\alpha_I\sim \Lambda^2\frac{(\sigma_X^{\rm SM})^{1/2}}{(\sigma^{\rm BSM}_{X,II})^{1/2}} $. These are $n$ points. 
\item[4)]
For every pair $(I,J)$, a single non-zero pair $(\alpha_I,\alpha_J)$  satisfying   $\alpha_{I,J}\gg1$ and $\alpha_I \sigma^{\rm BSM}_{X,II}\sim \alpha_J \sigma^{\rm BSM}_{X,JJ}$. These are $n(n-1)/2$ points.
\end{enumerate}

We thus end up with a system of $(n+1)(n+2)/2$ linear equations that has to be solved. A useful side effect of the optimization described in the previous subsection is that the calculation of the various components becomes more transparent. In particular, 1) provides $\sigma_X^{\rm SM}$ and 2) provides the $\sigma^{\rm BSM}_{X,II}$ components. 3) provides the interference terms $\sigma^{\rm int}_{X,I}$ which can be simply obtained by subtracting $\sigma_X^{\rm SM}+\alpha_I^2 \Lambda^{-4} \sigma^{\rm BSM}_{X,II}$ from the outcome of the simulations in 3), instead of using the exact formulas of Eqs.~\eqref{eq:sys}. This  simplification is a consequence of having used arbitrary large $\alpha_I$'s in 2). Similarly, 4) provides the $\sigma^{\rm BSM}_{X,IJ}$ components, which are obtained by subtracting $\alpha_I^2\sigma^{\rm BSM}_{X,II}+\alpha_J^2\sigma^{\rm BSM}_{X,JJ}$ from the outcome of the simulations in 4). An analysis similar to the one of section~\ref{se:Opt} shows that the uncertainty on the $\sigma^{\rm BSM}_{X,IJ}$ terms is minimized for $\alpha_I \sigma^{\rm BSM}_{X,II}\sim \alpha_J \sigma^{\rm BSM}_{X,JJ}$.

\paragraph{Comments on Monte Carlo simulations}

Differential distributions are often estimated using Monte Carlo simulations, which reproduce the experimental setup assuming a fixed number of events $N_{\rm MC}$ or a fixed integrated luminosity ${\mathscr{L}}_{\rm MC}$. The expected, relative uncertainties associated with estimation of the event rate is given by $\sqrt{\bar V} \equiv 1/\sqrt{N_{\rm MC}}(1+O(N_{\rm MC}^{-1}))$ in both cases (see App.~\ref{app:MC}).


In order to search for a small deviation in a given set of data, one should require that the MC error be small with respect to the statistical error of the data. 
For binned data, the number of events in each bin $\hat N_{r}$ is Poisson-distributed and its relative statistical error is given by $1/\sqrt{\hat N_{r}}$. 
The requirement that the MC error be small with respect to the experimental error in every bin translates into the condition
\be
N_{{\rm MC},r}\gg \hat{N}_r\,\quad \textrm{for any bin } r\,. \label{eq:cond_MC_actual}
\ee
Note that this condition depends on the actual data one wants to analyse. It applies for both background-only  and signal  hypothesis, \ie~for both $\alpha= 0$ and $\alpha\neq 0$. 
Finally, the binning and range of all the MC histograms have to match the bins chosen for the data and are thus completely fixed.


\section{A concrete example: search for anomalous trilinear gauge coupling at the LHC}\label{se:searches}
To validate our reconstruction method, we apply it to the concrete example of the search  
for the dimension-6 effective operator $\mathcal{O}_{3W}$ in $WW$ production at LHC. The operator $\mathcal{O}_{3W}$ is defined as
\be
\mathcal{O}_{3W}=\varepsilon_{ijk} W^{i}{}_{\mu\nu} W^{j,\nu}{}_{\rho} W^{k,\rho\mu}\,
\ee
and its coefficient is denoted by $\frac{\alpha_{3W}}{\Lambda^2}$.

After electroweak symmetry breaking, it contributes to anomalous triple gauge couplings \cite{Hagiwara:1986vm} that can be parametrized as follows~\footnote{We follow the conventions of \cite{Fichet:2013ola}} 
\be
\mathcal{L}_{\rm CGC}^{\partial}=\lambda^Z\left[ig_Z Z_{\mu\nu} (\hat W^-_{\nu\rho}\hat W^+_{\rho\mu}-\hat W^+_{\nu\rho}\hat W^-_{\rho\mu})\right]
+\lambda^\gamma\left[ie F_{\mu\nu}(\hat W^-_{\nu\rho}\hat W^+_{\rho\mu}-\hat W^+_{\nu\rho}\hat W^-_{\rho\mu})\right]\,,
\ee
where $\hat W^+_{\mu\nu}=D_\mu W_\nu^+-D_\nu W_\mu^+$ and
\be\lambda^Z=\lambda^\gamma=3\frac{\alpha_{3W}}{g\,\Lambda^2}\,.\ee
The lagrangian $\mathcal{L}_{\rm CGC}^{\partial}$ induces new vertices among the weak gauge bosons which carry extra derivatives with respect to the Standard Model ones.
This new interactions will potentially deform the $WW$ differential rates at the LHC, especially in the high energy range of the distributions.

A search for the ${\cal O}_{3W}$ operator has been performed by the CMS collaboration in~\cite{CMS_search} where they consider $W^+W^-$ production in the leptonic decay channel at the LHC, with an integrated luminosity of $19.4$ fb$^{-1}$ at $8$ TeV center-of-mass energy. In~\cite{CMS_search} the same operator is defined using a different normalization convention and the translation to our notation is done by using the following relation:
\be \label{eq:transl}
\alpha_{3W}=\frac{g^3}{4}c_{WWW}\,.
\ee
Our aims are: \begin{itemize}
\item Determining the deformations of the differential rates in $W^+W^-$ production induced by the ${\cal O}_{3W}$ operator using our optimized technique for Monte Carlo simulations,
\item Deriving a 95\% CL bound on $\frac{\alpha_{3W}}{\Lambda^2}$ using the measured differential distributions of~\cite{CMS_search}.
\end{itemize}
Therefore we consider the process $p p \to W^+ W^-\to l^+ \nu l^- \bar\nu$ ($l=e,\mu$) at $8$ TeV center-of-mass energy.
\begin{figure}
\begin{picture}(400,120)
\put(0,-10){        \includegraphics[trim=22cm 3cm 14cm 3cm, clip=true,width=5cm]{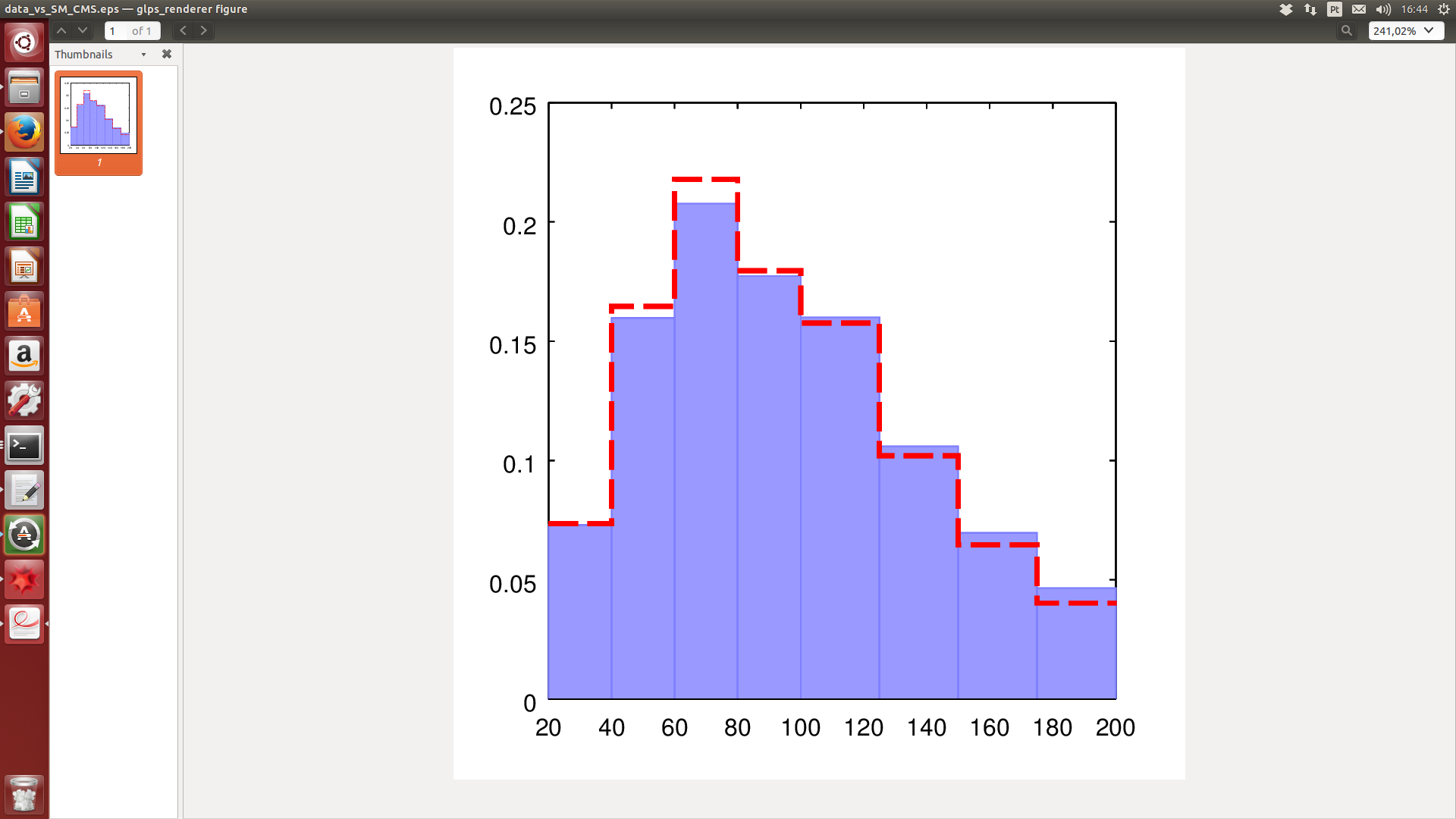}}
\put(153,-10){        \includegraphics[trim=22cm 2cm 14cm 5cm, clip=true,width=5cm]{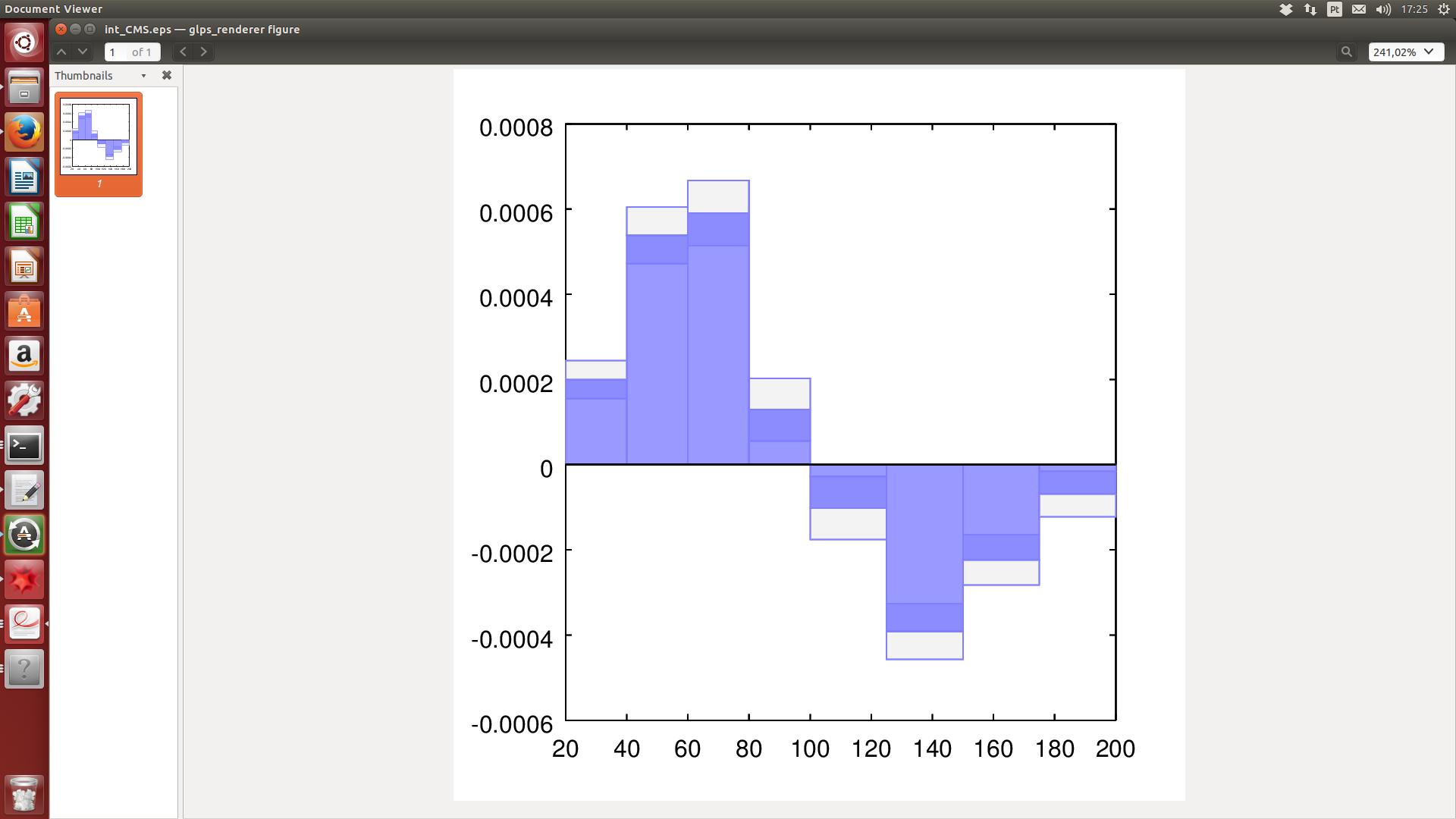}}
\put(305,-10){        \includegraphics[trim=21.5cm 3cm 14cm 3cm, clip=true,width=5.1cm]{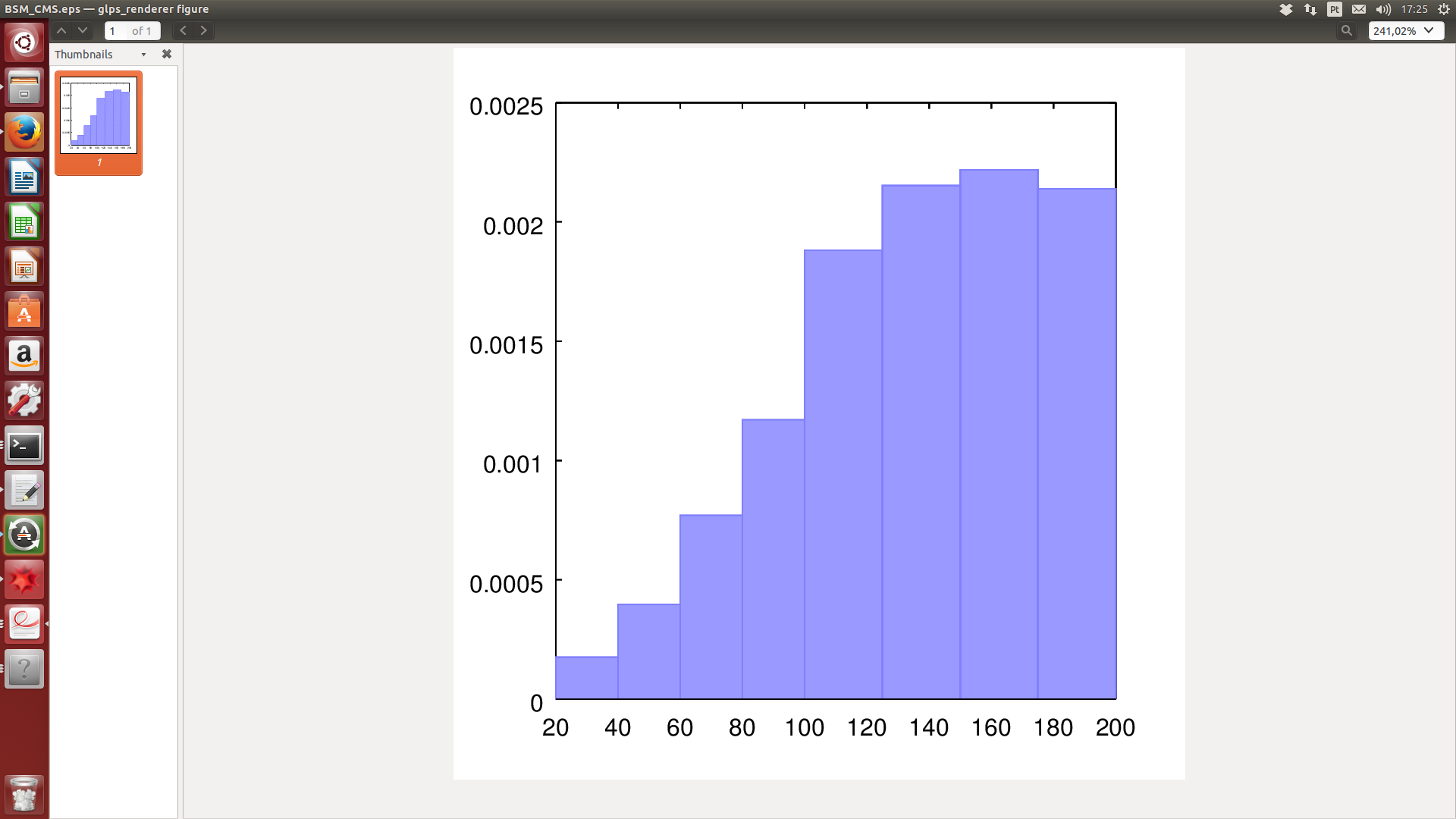}}
\put(70,-15){$\Scale[0.8]{m_{ll}}$}
\put(220,-15){$\Scale[0.8]{m_{ll}}$}
\put(370,-15){$\Scale[0.8]{m_{ll}}$}
\put( -5,30){\rotatebox{90}{$ \Scale[0.7]{\sigma_{\rm SM}^{-1}\,d\sigma^{\rm SM}/dm_{ll}}$ }}
\put(146,30){\rotatebox{90}{$ \Scale[0.7]{\sigma_{\rm SM}^{-1}\,d\sigma^{\rm int}/dm_{ll}~[{\rm TeV}^2]}$ }}
\put(300,30){\rotatebox{90}{$ \Scale[0.7]{\sigma_{\rm SM}^{-1}\,d\sigma^{\rm BSM}/dm_{ll}~[{\rm TeV}^4]}$ }}
\end{picture}
\vspace{0.5cm}
    \caption{   
     \label{fig:comp} Differential distributions of dilepton invariant mass. The $\sigma^{\rm SM}_{m_{ll}}$, $\sigma^{\rm int}_{m_{ll}}$, $\sigma^{\rm BSM}_{m_{ll}}$ components normalised to $\sigma^{\rm SM}$ are shown from left to right. The differential distribution from CMS data is shown in red. }
    \end{figure}
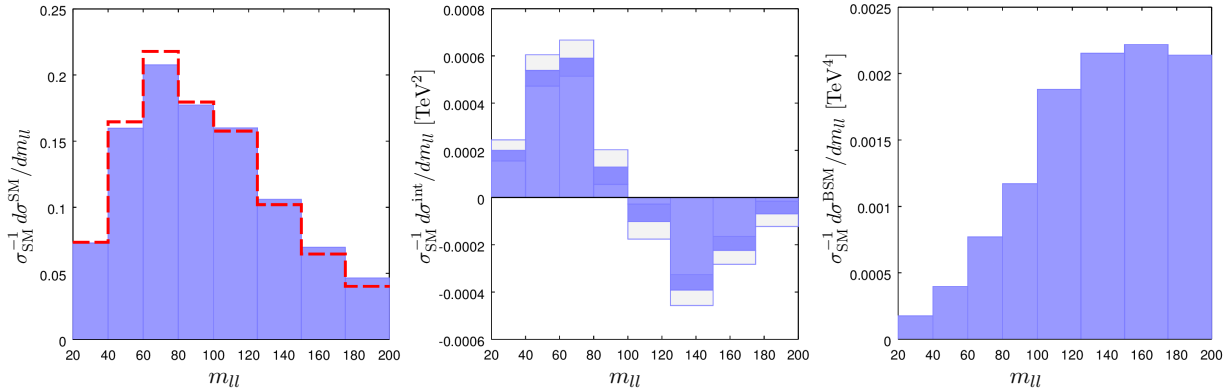
The measured unfolded differential distributions for this process are displayed in Fig.~3 of \cite{CMS_search} and they can be consistently compared to the outcome of our MC simulations. The dilepton invariant mass ($m_{ll}$) distribution in the ``$0$-jet category'' is the one chosen to put a bound on $\alpha_{3W}/\Lambda^2$. The total number of $W^+W^-$ + background events measured in the $0$-jet category is $\sim 4800$ and the quoted 95\%~CL bound~\cite{CMS_search} translated to our notation is
\be
-0.39<\frac{\alpha_{3W}}{\Lambda^2}<0.41 \,\,\textrm{TeV}^{-2}\,.\label{eq:CMS_bound}
\ee
We simulate events for $W^+W^-$ production at $8$~TeV with \texttt{MadGraph5}~\cite{Alwall:2011uj} after having implemented the ${\cal O}_{3W}$ perator in \texttt{FeynRules2.0}~\cite{Alloul:2013bka}. These events are showered with \texttt{Pythia8}~\cite{Sjostrand:2006za} and selected using the cuts chosen in the CMS analysis. In particular, the leptons are required to have $p_T>20$~GeV and $|\eta|<2.5$. Events with one or more jets with $p_T>30$~GeV and $|\eta|<4.7$ are rejected.

Following the notation introduced in the previous sections, the differential rate along the $m_{ll}$ variable will be denoted by $\sigma_{m_{ll}}$. 
We consider the binned distributions for $\sigma_{m_{ll}}$ in the range $[20,200]$ GeV, accordingly to the choice made in the CMS analysis.

We evaluate the components of the binned $m_{ll}$ distribution following our optimal method described in section~\ref{se:shape_analysis_EFT}.
We first compute the SM component $\sigma_{m_{ll}}^{\rm SM}$ by setting $\alpha_{3W}=0$ in our Monte Carlo simulation. Then we compute the $m_{ll}$ distribution for a very large value of $\alpha_{3W}/\Lambda^2$, chosen to be ~$272$~TeV$^{-2}$, which turns out to be proportional to the BSM component $\sigma_{m_{ll}}^{\rm BSM}$ to a very good approximation. The binned $\sigma_{m_{ll}}^{\rm SM}$ and $\sigma_{m_{ll}}^{\rm BSM}$ components are shown in Fig.~\ref{fig:comp}.

In order to compute the interference component $\sigma_{m_{ll}}^{\rm int}$ following the results of section~\ref{se:Opt}, we have used Eq.~\eqref{eq:alpha_opt} to determine the third optimal value of $\alpha_{3W}/\Lambda^2$.  The  $\sigma^{\rm SM}/\sigma^{\rm BSM}$ ratio being  roughly about $100$ TeV$^{-4}$ in most of the bins, we conclude that the third optimal point for the simulation is roughly $\alpha_{3W}/\Lambda^2 \approx 10 $~TeV$^{-2}$.  

\begin{figure}
\begin{picture}(400,155)
\put(80,0){        \includegraphics[trim=18cm 4cm 10cm 8cm, clip=true,width=8cm]{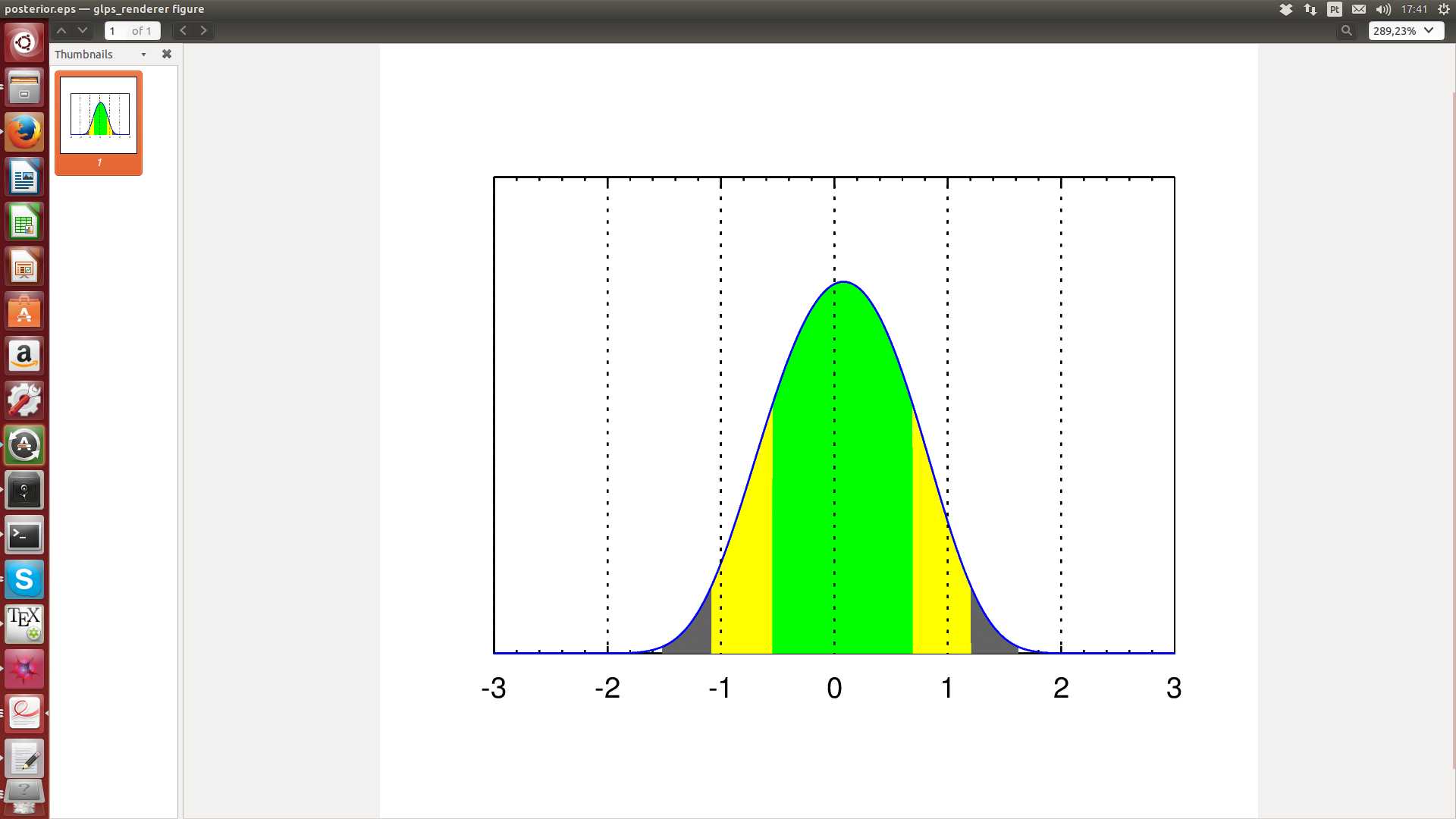}}
\put(180,-5){$\alpha_{3W}/\Lambda^2$ [TeV$^{-2}$]}
\end{picture}
\vspace{0.5cm}
    \caption{   
     \label{fig:8TeV_posterior}
Posterior density probability for the  coefficient of ${\cal O}_{3W}$.
Green, yellow and grey areas  correspond to Bayesian credible regions with respectively
$68.27\% \, , \, 95.45\% \, , \, 99.73\% $ probability.
     }
    \end{figure}

We still need to set the size of our MC samples. We require, for every bin, the number of MC events to be larger than the observed data analyzed by the CMS collaboration in the search for the ${\cal O}_{3W}$ operator. For example, $N_{\rm MC}=5\cdot 10^5$ events would give a MC uncertainty that is typically ten times smaller than the statistical ones. Although this number is enough for the sake of analyzing the CMS distribution, it turns out that this amount is not sufficient to resolve the binned interference components because the interference is further suppressed by a $m_W^2/E^2$ factor with respect to the naive expectation. This can be understood thanks to some helicity selection rules~\cite{Azatov:2016sqh,Falkowski:2016cxu}.

It turns out that a much larger number of events is needed to properly estimate the interference component. We have indeed used $N_{\rm MC}=2.4\cdot 10^6$ events for each of the three simulation points $\alpha_{3W}/\Lambda^{2}=0,\,8.6$, and $272$~TeV$^{-2}$. The relative covariance matrix calculated in section~\ref{se:Opt} readily provides the uncertainty on the $\sigma_{m_{ll}}^{\rm SM}$, $\sigma_{m_{ll}}^{\rm int}$ and $\sigma_{m_{ll}}^{\rm BSM}$ components. 

The uncertainties on $\sigma_{m_{ll}}^{\rm SM}$ and $\sigma_{m_{ll}}^{\rm BSM}$ are too small to be visible in Fig.~\ref{fig:comp}. In contrast, the uncertainty on $\sigma_{m_{ll}}^{\rm int}$ is not negligible, even with such a large MC sample. Having fixed a unique optimized value of $\alpha_1$ for every bin and taking into account that the number of MC events in a given bin depends on this $\alpha_1$, the uncertainty on $\sigma^{\rm int}_r$ is obtained from Eq.~\eqref{eq:Cov_genV}, where the relative variances associated to a given bin is 
\be 
\bar V^i_r= \frac{\sigma^i_{\rm tot}}{\sigma^i_r}\frac{1}{N_{\rm MC}} \,.
\ee
These uncertainties are shown in Fig.~\ref{fig:comp}.

Finally, we compare our reconstructed differential rate for $pp\rightarrow W^+W^-\rightarrow l^+ \nu l^-\bar\nu$ to the measured one shown in Fig.~3 of the CMS study~\cite{CMS_search}. The uncertainties quoted are a combination of statistical and systematic errors. Because combinations of many sources of uncertainties tend to be governed by the central limit theorem \cite{Fichet:2016gvx}, we approximate the likelihood for each bin as a Gaussian. 
The complete likelihood used in our analysis is 
 \be
L(\alpha_{3W})=\prod_r \, \exp\left[-  \frac{1}{  2\,\Delta_r^2}\left( \frac{\sigma_r(\alpha_{3W})-\sigma_r^{\rm obs}}{\sigma_r^{\rm SM}} \right)^2  \right]\,,
\label{eq:like}
 \ee
where  $\sigma_r^{\rm obs}/\sigma_r^{SM}$ are the experimental numbers given in \cite{CMS_search} and $\Delta_r$ are the combined uncertainties for each bins, which are typically $\sim 8\%$.
Using Eq.~\eqref{eq:like}, we compute the credible  intervals for the $\alpha_{3W}/\Lambda^2$ parameter, assuming a flat prior over $[-100,100]$~TeV. 
The posterior distribution for $\alpha_{3W}\Lambda^{-2}$ is shown in Fig.~\ref{fig:8TeV_posterior}.  We find
\be
-1.09<\frac{\alpha_{3W}}{\Lambda^2}<1.19 \,\,\textrm{TeV}^{-2} \quad\,\textrm{at 95\% CL}\,\,. \label{eq:US_bound}
\ee
This bound is in agreement within less than two sigma with the one reported by CMS (see Eq.~\eqref{eq:CMS_bound}).

\section{Conclusion}\label{se:conc}

If new particles beyond the Standard Model are too heavy to be on-shell produced at the LHC, their presence can still be indirectly detected via the effect of SM higher dimensional operators. In such scenario, the LHC precision physics program would play a central role for new physics searches. 
The key observables for revealing the existence of higher dimensional operators may be the  distributions of final state kinematic variables, that contain precious information about new physics effects.
The analysis of these differential rates thus deserves to be optimized in all its aspects. 

We focus on the case were leading effects from new physics arise from dimension-6 operators. 
We first inspect the event rates  and the problem of detecting the deformations induced by the presence of the effective operators.  
We make clear that, in general, the   pure BSM term should not be neglected in the differential rate analysis--even though it is $O(\alpha^2/\Lambda^{4})$, since there can be regions of  phase space where its contribution is dominant. We have also found a bound on the interference term which follows from the application of the Cauchy-Schwartz inequality.  Using this bound, it can be quantitatively shown that regions of phase space with rare SM events are very  important to search for deformations of the differential rates.

Based on this preliminary analysis, we determine an optimal method to obtain the different contributions to the differential rates in the presence of dimension-6 effective operators, assuming that an estimator (\textit{e.g.} a Monte Carlo tool) of the distributions is available. 
In the case of $n$ effective operators, the evaluation of the rate at $(n+1)(n+2)/2$ different points are needed. The various contributions to the differential rate are then simply obtained by solving a linear system. 

A crucial aspect of the proposed method is the minimization of the uncertainty through an optimal choice of the higher dimensional operators coefficient to be used in the simulations. 
In the case of a single dimension-6 operator we have to estimate the differential rate for three values of the coefficient $\alpha$. The analysis of the relative covariance matrix of the three estimators reveals that the uncertainty is minimized for values of $\alpha$ equal to zero, infinity, and
$\pm\Lambda^{2}\sqrt{\sigma_X^{\rm SM}/\sigma_X^{\rm BSM}}$. Interestingly, this result turns out to be independent of the value of the interference.
The covariance matrix  provides the uncertainties on the estimated contributions and their correlations, allowing a well-defined control of the estimations. This covariance matrix should be in principle implemented in any subsequent statistical analysis. It turns out that the uncertainty on the interference component is larger than the ones on the SM and BSM components by a factor $3/\sqrt{2}$ if the interference saturates the positive Cauchy-Schwartz bound, and grows very quickly if the interference is smaller than this value. 

We illustrate and check our method by determining the deformations induced by the ${\cal O}_{3W}$ operator on leptonic  final states from $WW$ production, at the $8$~TeV LHC. We work at reconstruction level, aiming to approximately reproduce
the analysis made by the CMS Collaboration in Ref.~\cite{CMS_search}.  We ultimately  reproduce within a two sigma range  the bound on $\alpha_{3W}/\Lambda^2$  obtained in this CMS analysis.

\section*{Note added}
After completing our work we became aware of similar developments made by members of the Higgs Cross Section Working Group ~\cite{Brenner:2143180} and the ATLAS collaboration \cite{ATL-PHYS-PUB-2015-047}. The basic method presented in these references (called {\it morphing}) is essentially the same as our reconstruction method described in Section~\ref{se:recon}, although the parametrization that has been used is slightly different. However, the question of how the input parameters need to be chosen such that the expected uncertainty of the output is minimal has not been addressed in these studies. Our paper fills this important gap by presenting, for the first time, an optimal morphing method  and the statistical approach which gives rise to it.

\section*{Acknowledgements} 

We thank Alicia Calder\'on and Rafael Coelho Lopes de S\'a for providing us invaluable information about the WW shape analysis done in CMS, and Alexandra Carvalho for collaborating with us in an earlier version of this work. 
We also would like to thank Rogerio Rosenfeld for useful discussions. 
The work of A.T. and S.F. was supported by the S\~ao Paulo Research Fundation (FAPESP) under grants 2011/11973-4,  2013/02404-1, and 2014/21477-2.
P.R.T. was supported by Brazilian Science without borders program from CAPES funding agency under grant BEX 11767-13-8.
\\
\\
\\

\noindent{\Large\bf Appendix}

\appendix
 
\section{Proof of  Eq.\eqref{eq:res}} \label{app:proof_Z}

For our purpose it is enough to assume a p-value based discovery test, that leads to Eq.~\eqref{eq:Z_disc}.
 A similar demonstration can be done using a Bayes factor.  Using the inequalitie in Eq.~\eqref{eq:ass} and keeping only the leading terms in the discovery tests, one gets
\be
Z_1\propto \frac{\alpha}{\Lambda}\frac{\sigma^{\rm int}_{1}}{\sqrt{\sigma^{\rm SM}_{ 1 }}}\,,\quad
Z_2\propto \frac{\alpha^2}{\Lambda^2}\frac{\sigma^{\rm BSM}_{2}}{\sqrt{\sigma^{\rm SM}_{ 2 }}}
\ee
Let us consider the ratio
\be
\frac{Z_1}{Z_2}=\frac{\Lambda}{\alpha} \frac{\,\sigma^{\rm int}_{1} \sqrt{\sigma^{\rm SM}_{2}}}{ \sqrt{\sigma^{\rm SM}_{1}}\sigma^{\rm BSM}_{2}}
\ee
Using the Cauchy-Schwartz bound on $\sigma^{\rm int}_1$ (see Eq.~\eqref{eq:int_bound}), one has 
\be
\frac{Z_1}{Z_2}\leq \frac{\Lambda}{\alpha} \frac{ 2\sqrt{\sigma^{\rm SM}_{2}\,\sigma^{\rm BSM}_{1}}}{ \sigma^{\rm BSM}_{2}}
\ee
Using  the equality of BSM components made in the initial assumptions (see Eq.~\eqref{eq:ass}), it comes
\be
\frac{Z_1}{Z_2} \ll 2\left(\frac{\Lambda^2}{\alpha^2}\frac{ \sigma^{\rm SM}_{2}}{ \sigma^{\rm BSM}_{2}}\right)^{1/2} \,.
\ee
The inequality contained in  Eq.~\eqref{eq:ass} then proves Eq.~\eqref{eq:res}.

\section{ Uncertainties on Monte Carlo estimators }
\label{app:MC}

For a fixed luminosity ${\mathscr{L}}_{\rm MC}$, the number of MC events $\hat N_{\rm MC}$ is Poisson-distributed. The estimator  of the event rate is given by \be\hat\sigma=\frac{\hat N_{\rm MC}}{\mathscr{L}_{\rm MC}}\,,\ee which satisfies
\be
{\rm E}[\hat \sigma]=\sigma\qquad\mbox{and}\qquad{\rm V}[\hat \sigma]=\frac{\sigma }{\mathscr{L}_{\rm MC}}\,,
\ee
where $\sigma$ is the theoretical rate.
The expected relative error associated with the estimation of $\sigma$ is thus given by $1/\sqrt{\sigma \mathscr{L}_{\rm MC}}$.

Instead of a fixed  luminosity, one can require a fixed number of events $N_{\rm MC}$. In this case, the random variable is the MC luminosity $\hat {\mathscr{L}}_{\rm MC}$. 
The combination $\sigma \hat {\mathscr{L}}_{\rm MC}$ follows an Erlang distribution
\be
\sigma \frac{(\sigma \hat {\mathscr{L}}_{\rm MC})^{N_{\rm MC}-1}}{(N_{\rm MC}-1)!} e^{-\sigma\hat {\mathscr{L}}_{\rm MC}}\,,
\ee 
for which ${\rm E}[\sigma \hat{\mathscr{L}}_{\rm MC}]={\rm V}[\sigma\hat{\mathscr{L}}_{\rm MC}]=N_{\rm MC}$.
In this case the estimator of the event rate is given by
\be\frac{1}{\hat\sigma}=\frac{ \hat{\mathscr{L}}}{N_{\rm MC}}\,,\ee which satisfies
\be
{\rm E}\left[\frac{1}{\hat\sigma}\right]=\frac{1}{\sigma}\qquad\mbox{and}\qquad {\rm V}\left[\frac{1}{\hat\sigma}\right]=\frac{1 }{\sigma^2}\frac{1}{ N_{\rm MC}}\,.
\ee
The expected relative error associated with the estimation of $1/\sigma$ is therefore given by $1/ \sqrt{N_{\rm MC}}$. For large $N_{\rm MC}$, one has that ${\rm E}[\hat \sigma] = E[1/\hat \sigma]^{-1}(1+O(N_{\rm MC}^{-1}))$ and the relative uncertainty associated with the estimation of $\sigma$ is given by $ 1/\sqrt{N_{\rm MC}}$ up to  $O(1/N_{\rm MC})$ corrections.

\section{General covariance matrix}
\label{app:cov}

The relative covariance matrix for the estimators $(\hat \sigma_r^{\rm SM}$, $\hat \sigma_r^{\rm int}$, $\hat \sigma_r^{\rm BSM})$, assuming arbitrary relative covariance for the $\hat\sigma_r^i$, namely $\bar V^0_r\neq \bar V^1_r \neq \bar V^2_r$, is given by 
\be
\begin{split} & \bar C_r(0,\alpha_1,\infty) =
\\
& \begin{pmatrix}
\bar V_r^0  & - \frac{\Lambda^2}{\alpha_1}\frac{\sigma_r^{\rm BSM}}{\sigma^{\rm int}_r}\,\bar V_r^0 & 0 \\
 -\frac{\Lambda^2}{\alpha_1}\frac{\sigma_r^{\rm SM}}{\sigma^{\rm int}_r}\bar V_r^0 &
 (\bar C_r)_{22}
  & -\frac{\alpha_1}{\Lambda^2}\frac{\sigma_r^{\rm BSM}}{\sigma^{\rm int}_r}\bar V_r^2 \\
0 & -\frac{\alpha_1}{\Lambda^2}\frac{\sigma_r^{\rm BSM}}{\sigma^{\rm int}_r}\bar V_r^2  & \bar V_r^2 
\end{pmatrix}\,, \label{eq:cov_full_gen}
\end{split}
\ee
where
\be
\begin{split}
 (\bar C_r)_{22}= 
\bar V_r^1\left(1+2\frac{ \Lambda^2\,\left(
\sigma^{\rm SM}_r+\alpha_1^2 \Lambda^{-4} \sigma_r^{\rm BSM}
\right)}{\alpha_1 \sigma_r^{\rm int}}+2\frac{  \sigma^{\rm SM}_r \sigma_r^{\rm BSM}}{ (\sigma_r^{\rm int})^2}\right)
\\+(\bar V_r^0+\bar V_r^1)\frac{ \Lambda^4\, (\sigma^{\rm SM}_r)^2}{\alpha_1^2 (\sigma_r^{\rm int})^2}
+(\bar V_r^1+\bar V_r^2)\frac{ \Lambda^{-4} \alpha_1^2\, (\sigma^{\rm BSM}_r)^2}{(\sigma_r^{\rm int})^2}\,.
\end{split} \label{eq:Cov_genV}
\ee

\bibliographystyle{JHEP} 

\bibliography{shape_biblio}

\end{document}